\documentclass[a4paper]{jpconf}
\usepackage{graphicx}
\begin{document}
\title{Finite size effects in hadron-quark phase transition \\ {\bf by the} Dyson-Schwinger method}

\author{Nobutoshi Yasutake$^1$, Huan Chen$^2$, Toshiki Maruyama$^3$, \\  Toshitaka Tatsumi$^4$}
\address{$^1$ Department of Physics, Chiba Institute of Technology,\\ 2-1-1 Shibazono, Narashino, Chiba 275-0023, Japan}
\address{$^2$ School of Mathematics and Physics, China University of Geoscience, Wuhan 430074, China} 
\address{$^3$ Advanced Science Research Center, Japan Atomic Energy Agency, \\Tokai, Ibaraki 319-1195, Japan} 
\address{$^4$ Department of Physics, Kyoto University, Kyoto 606-8502, Japan}
\ead{nobutoshi.yasutake@p.chibakoudai.jp}

\begin{abstract}
We study the hadron-quark phase transition, taking into account the finite-size effects for neutron star matter.
For the hadron phase, we adopt a realistic equation of state within the framework of the Brueckner-Hartree-Fock theory. For the quark phase, we apply the Dyson-Schwinger method. The properties of the mixed phase are clarified by considering the finite-size effects. We find that, if the surface tension is strong enough, the equation of state becomes to be close the one with the Maxwell condition, though we properly adopt the Gibbs conditions. 
This result is qualitatively the same with the one by the use of the simple bag model. We also find that the mass-radius relation by the EoS is consistent with the observations of massive neutron stars.
\end{abstract}

\section{Introduction}

The equation of sate~(EoS) is one of the most important ingredient for the study of compact stars. It is also related to the QCD phase diagram, which has a large uncertainty. The recent observation of Ref.~\cite{demorest10} strongly restricts the phenomenological EoS; the EoS must explain $\sim 2 M_\odot$. On the other hand, the statistical approach from many observations~\cite{steiner12} gives us prospect of fixing of the EoS. 
Theoretically, the study on the baryon-baryon interaction has been recently a hot topic from lattice QCD~(LQCD) simulations ~\cite{ishii07}. Experiments such as JPARC will also provide useful information about the baryon-baryon interaction in the future. Hence, we should not neglect their impacts on the EoS: the EoS in the hadron phase may be provided by using the baryon-baryon interactions thus obtained in near future. In this paper we use the result by the Brueckner-Hartree-Fock~(BHF) theory for hadron matter, which is based on the baryon-baryon interactions~\cite{baldo99, schulze95}.

For the quark phase, there are a lot of the effective models of QCD, such as the bag model, the (P)NJL model, and so on. However, most of them can not sufficiently describe the realistic features of QCD. For example, the mass functions of quarks should have the momentum dependence as the recent studies by LQCD show~\cite{parappilly06}. Here, we introduce the Dyson-Schwinger(DS) method which shows the momentum dependence on the mass function. This model is not based on the mean-field approximation. Instead, we need to calculate the quark propagators directly~\cite{huan11}.

The main purpose of this study is to know the property of the mixed phase between the quark phase and the hadron phase considering {\it the finite-size effects}. Assuming the quark deconfinement transition to be of first order, we needs the careful treatment of the effects in the phase transition of the multi-component system. 
Generally the properties of the mixed phase should strongly  depend on the electromagnetic interaction and surface tension, and these  effects, sometimes called ``{\it the finite-size effects}", lead to the non-uniform ``Pasta" structures~\cite{dma02}. The charge screening is also important for their mechanical instability. 
In the previous papers these finite-size effects have been properly taken into account to elucidate the properties of the pasta structure and demonstrate the importance of the charge screening adopting a simple bag model~\cite{maruyama07, yasutake09}. 
However, the adopted quark model, the bag model, is too simple, and we introduce the DS model for it as mentioned in the last paragraph. 
Note that the mass-radius relation by our EoS is consistent with the both observations, Ref.~\cite{demorest10} and Ref.~\cite{steiner12}. 

This article is organized as follows. In Sec.~II, we outline our framework. In Sec.~III, we present numerical results. Sec.~IV is devoted to the discussion.

\section{Equation of State and finite-size effects}

\subsection{Dyson-Schwinger method}
For the deconfined quark phase, we adopt a model based on the DS equation of QCD, which provides a continuum approach to QCD that can simultaneously address both confinement and dynamical chiral symmetry breaking. It has been applied with success to hadron physics in vacuum, and to QCD at nonzero chemical potential and temperature. The details can be found in Ref.\cite{huan11}. Here, we just show brief points for this article.

Unfortunately, it is not possible to solve the DS equations precisely, and we introduce  some approximations; e.g. for the quark-gluon vertex, we use the ``rainbow approximation"\cite{bender96}. 
For the gluon propagator in the vacuum, we assume a Gaussian-type effective
interaction $\mathcal{G}(k^2; \mu)$\cite{alkofer02} in the Landau gauge
\begin{equation}
{\mathcal G}(k^2; \mu)=
\frac{4\pi^2D}{\omega^6} {\rm e}^{-\alpha\mu^2/\omega^2} k^2\,
{\rm e}^{-k^2/\omega^2} \:,
\label{IRGsmu}
\end{equation}
with the parameter $\alpha$ we introduced to quantify the asymptotic freedom at large chemical potential $\mu$. At the extreme limit of $\alpha \approx \infty$, the model is reduced to a free quark system at finite chemical potential. Here, we take two typical values $\alpha=1$ and $\alpha=2$ in our calculation as in Ref.~\cite{huan11}, and choose the other set of parameters $\omega=0.5\;{\rm GeV}$, $D=1\;{\rm GeV}^2$ \cite{alkofer02}.
In the following we will study the parameter dependence of our results, marked as ``DS$\alpha$". We consider three quark flavors $q=u,d,s$,
which are independent of each other in our model. We take the current-quark masses $m_{u,d}=0$ for simplicity and $m_s=115\;{\rm MeV}$  \cite{alkofer02}. 
From the quark propagator, one can calculate the quark number density
and correspondingly obtain the physical quantity such as pressure~\cite{huan11}.

\subsection{Brueckner-Hartree-Fock theory}
Our theoretical framework for the hadron phase of matter is the nonrelativistic BHF approximation based on the microscopic nucleon-nucleon potentials. The BHF calculation is a reliable and well-controlled theoretical approach for the study of dense baryonic matter. Detailed procedure can be found in Refs~\cite{baldo99, schulze95}. 
In this study, we adopt the result with one of the Bonn potentials~\cite{machleidt87} for the nucleon-nucleon interacti, which is called Bonn-B~(BOB) potential. Here we do not consider the degree of freedom of hyperons, since they are reduced by the existence of quarks as we suggested~\cite{maruyama07}. Of course we should show the behavior including hyperons, and we leave it in near future.

\subsection{Mixed phase and finite-size effects}
To take into account the finite-size effects, we impose the Gibbs conditions on the mixed phase, which require the pressure balance and the equality of the chemical potentials between two phases besides the thermal equilibrium. We employ the Wigner-Seitz approximation in which the whole space is divided into equivalent cells with given geometrical symmetry, specified by the dimensionality  $d=3$ (droplet or bubble), $d=2$ (rod or tube), or $d=1$ (slab). The structures of tube and bubble are characterized by the reversed particle distributions of rod and droplet.  A sharp boundary is assumed between two phases and the surface energy is taken into account in terms of a surface-tension parameter $\sigma$. The surface tension of the hadron-quark interface is poorly known, but some theoretical estimates (e.g. lattice gauge simulations at finite temperature \cite{kajantie91}) suggest a range of $\sigma \approx 10$--$100\;\rm MeV\! ~fm^{-2}$. We show results using $\sigma=40\;\rm MeV\! ~fm^{-2}$ and $\sigma=80\;\rm MeV\! ~fm^{-2}$ in this article. The general discussions about the mixed phase in quark-hadron phase transition are given in our recent review~\cite{yasutake13}.

\section{Results}
\begin{figure}[h]
\includegraphics[width=19pc]{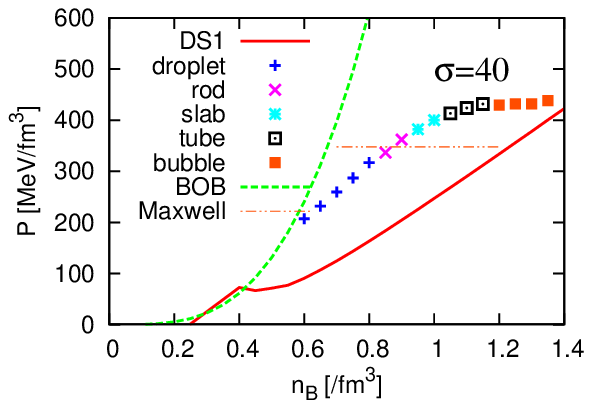}
\includegraphics[width=19pc]{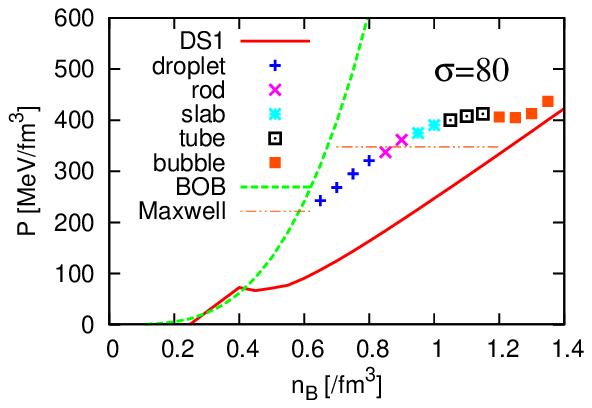}
\caption{\label{label} EoS with the quark model `` DS1", where the energy of the surface tension $\sigma$ is $\sigma = 40$ MeV fm$^{-2}$ for the left panel, and $\sigma = 80$ MeV fm$^{-2}$ for the right one. Dashed line shows the one for  the hadron EoS, which is labeled with gBOBh. Solid curves show the quark model gDS1h.  Each dot shows the mixed phase with the non-uniform structures. Narrow dashed-solid lines show the phase transition under the Maxwell construction. }
\label{fig:01}
\end{figure}

In Figure~\ref{fig:01}, we show the EoS based on the above descriptions with ``DS1" as one example.
The left panel show the case with $\sigma=40\;\rm MeV\!~fm^{-2}$, and the right one is for $\sigma=80\;\rm MeV\! fm^{-2}$. Dashed lines show the pure hadron matter, and the solid lines do the pure quark matter.
The EoS of the mixed phase becomes similar to the one under the bulk Gibbs calculation for weak surface tension, and to the one given by the Maxwell construction for strong surface tension. This behavior is universal, so that this result is qualitatively the same with the one with simple bag model~\cite{maruyama07}. 

\begin{figure}[h]
\includegraphics[width=27pc]{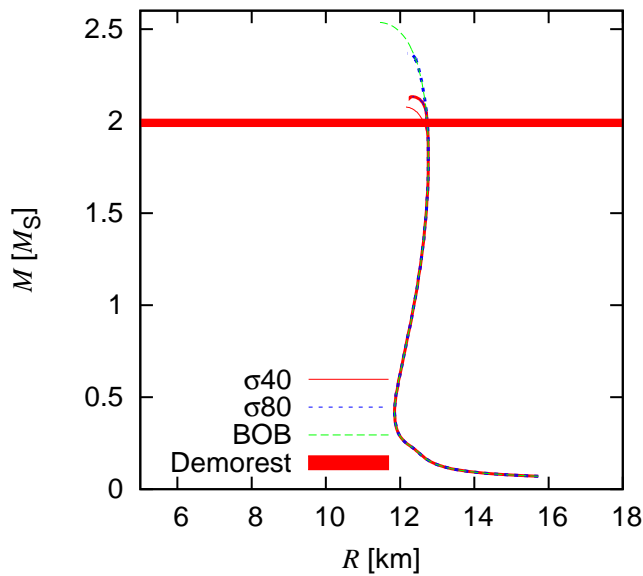}
\begin{minipage}[b]{11pc}\caption{\label{label} Mass-radius relations for EoS's shown in Figure~\ref{fig:01}. The solid lines (short-dashed lines) show those with the mixed phase, where the surface tension is $\sigma=40~(80)$ MeV fm$^{-2}$.  Dashed line shows the one for hadron EoS, which is labeled with ``BOB". Thick curves show the quark model ``DS1", and narrow curves ``DS2". Here, we also show the observational result by Demorest et al.; $M$ = (1.97$\pm$ 0.04) $M_\odot$~\cite{demorest10}. }
\label{fig:02}
\end{minipage}
\end{figure}

At the last part of this section, we briefly discuss some implications of our results on the mass-radius relation of neutron stars shown in Fig.~\ref{fig:02}, which is obtained by solving the Tolman-Oppenheimer-Volkoff equation. For the crust, we use the BPS EoS which is widely used~\cite{bps}. 
Clearly we can see that the maximum masses for all our models are slightly larger than the recent observational data by Demorest et al.~\cite{demorest10}. 
Moreover, most of neutron stars by our models exist around $R \sim12$~km. They are consistent with the statistical approach based on the observations by Steiner et al.\cite{steiner12}, though this constraint is not so strong compared with the one by Demorest et al.~\cite{demorest10}. 

\section{Discussion}
We have studied the hadron-quark phase transition with the finite-size effects for neutron star matter. For the hadron phase, we adopt a realistic equation of state in the framework of the BHF theory, we do the DS method for the quark phase. We find that, if the surface tension is strong, the EoS becomes to be close the one with the Maxwell condition. We also find that the mass-radius relation by the EoS is consistent with the observations of massive neutron stars.

\ack
{NY is grateful to H. J. Schulze, F. Burgio, M. Baldo, and D. Blaschke for their warm hospitality and fruitful discussions. This work was partially supported by the Grant-in-Aid for the Global COE Program gThe Next Generation of Physics, Spun from Universality and Emergenceh from the Ministry of Education, Culture, Sports, Science and Technology (MEXT) of Japan and the Grant-in-Aid for Scientific Re-search (C) (25105510, 23540325).}

\end{document}